\pdfoutput=1
\documentclass[a4paper,11pt]{article}
\usepackage[utf8x]{inputenc}
\usepackage[OT4]{fontenc}
\usepackage{graphicx}
\usepackage{amsmath}
\usepackage{amssymb}
\usepackage[nooneline]{subfigure}

\author{Ryszard Winiarczyk \and 
Piotr Gawron \and 
Jarosław Adam Miszczak \and 
Łukasz Pawela \and 
Zbigniew Puchała\\[1em]
Institute of Theoretical and Applied Informatics,\\
Polish Academy of Sciences\\
Bałtycka 5, 44-100 Gliwice, Poland
}

\date{11/12/2012}
\title{Analysis of patent activity in the field of quantum information processing}

\begin{document}
\maketitle

\begin{abstract}
This paper provides an analysis of patent activity in the field of quantum
information processing. Data from the PatentScope database from the years
1993-2011 was used. In order to predict the future trends in the number of filed
patents time series models were used.
\end{abstract}

\section{Introduction}
Quantum information science studies the application of quantum mechanics to
processing, storing and transmitting information. The main trends of research
involving theoretical aspects of quantum computing are: quantum information
theory,  quantum computation, quantum cryptography, quantum communication and
quantum games. The origins of the field date back to the first decades of the
twentieth century, when the bases of quantum mechanics were formulated.  The
beginnings of quantum information theory can be traced back to von Neumann
\cite{von1927thermodynamik}, but the first established research in this area was
conducted by Holevo \cite{holevo73information} and Ingarden
\cite{ingarden76quantum}.

In the last two decades of the twentieth century, the groundwork for quantum
computing was laid. First in the eighties, Bennett and Brassard
\cite{bennett1984quantum}, then seven years later Ekert \cite{ekert1991quantum} 
discovered quantum key distribution (QKD) protocols. These research 
achievements gave rise to the whole new field of quantum cryptography. The
Feynman's idea  of quantum simulators \cite{feynman1982simulating} and Deutsch's
work  \cite{deutsch1985quantum} on the universal quantum computer, followed by
the discovery of the first  quantum algorithms by Shor \cite{shor1994algorithms}
and Grover \cite{grover1996fast} in the nineties of the twentieth century
created the whole new field of research on quantum computing and quantum
algorithms. One of the promising areas of quantum information processing is 
quantum game theory initiated by Meyer \cite{meyer1999quantum} and then followed
by  Eisert, Wilkens and Lewenstein \cite{eisert1999quantum} at the turn of the
century. Its application to quantum auctions was proposed later by Piotrowski
and Sładkowski \cite{piotrowski2002quantum}.

At present it is hard to tell which specific physical system  will be applied
for the implementation of quantum information processing~\cite{hemmer09where}. 
Among the most promising physical implementations one can point out cavity
quantum  electrodynamics \cite{walther06cavity,miller05trapped,stute12toward},
trapped  ions \cite{kim09integrated,montz11-14qubit}, quantum optical lattices 
\cite{brennen99quantum,li08real-time,everitt11creating}, nitrogen-vacancy 
centers \cite{gurudev-dutt07quantum,maurer12room-temperature} and various 
realizations based on superconducting qubits \cite{jordan04entanglement,
chirolli06full,houck09life,gronbech-jensen10tomography,rigetti12superconducting,
stojanovic12quantum-control}. However, it is believed that realistic systems
capable of processing information according to the rules of quantum theory will
be composed of  diverse physical systems interacting with each
other~\cite{kimble08quantum}. For this reason a significant amount of research
has been dedicated to   engineering the building blocks of quantum
networks~\cite{clausen11storage,hall11ultrafast,choi11athome}.

\section{Patent analysis}
The main goal of this work is to analyse and forecast patent activity in the 
field of quantum information processing. In order to reach this goal a 
statistical analysis of the patent database was made.

\subsection{Research methodology}

\subsubsection{Data source}
In order to gather patents covering quantum information processing the database 
of patent applications PatentScope \cite{patentscope}, which is made accessible
by World Intellectual Property Organization, was used.

\subsubsection{Search methodology}
The PatentScope database allows searching in full text versions of patent
applications at European, international and country levels. It allows searching
in almost eleven million patent documents which includes 2 million of
international patent applications.

In order to analyse patent documents which may concern quantum information 
processing, following queries to PatentScope database were performed. 

\begin{enumerate}
    \item 
    \texttt{"quantum computer" OR "quantum computing" OR 
    "quantum computation" OR "quantum compute" 
    }
    \item 
    \texttt{"quantum communication"     
    }
    \item 
    \texttt{"quantum information"       
    }
    \item 
    \texttt{teleportation       
    }
    \item 
    \texttt{"quantum bit" OR qubit OR qbit      
    }
    \item 
    \texttt{quantum AND "random number generator"       
    }
    \item 
    \texttt{"quantum cryptography"      
    }
    \item 
    \texttt{"quantum key" AND (distribut* OR exchang*) 
    }
    \item 
    \texttt{"quantum Fourier"   
    }
    \item 
    \texttt{"quantum fast" OR "fast quantum"    
    }
    \item 
    \texttt{(quantum OR photo* OR optic*) AND BB84      
    }
    \item 
    \texttt{quantum AND grover  
    }
    \item 
    \texttt{quantum AND ("single photon source" OR "single-photon source" OR
    "single \\ photon generator" OR "single-photon generator" OR "single photon
    detector" \\ OR "single-photon detector") 
    }
    \item 
    \texttt{"quantum switch"    
    }
    \item 
    \texttt{quantum AND spintronic*     
    }
\end{enumerate}

Selection of those queries was based upon previous work presented in 
\cite{chang}. The PatentScope search engine was instructed to apply queries only
to the front  page of the patent text. The reason for this is that many patents
mention  quantum devices or qubits but they cannot be reasonably classified as 
concerning directly quantum information processing. The problem is discussed in
the next subsection.

\subsection{Analysis of patents concerning quantum computing in the years 1993-2012}
Analysis of patent data is one of the most reputed methods of projecting 
technology development. Arguments supporting this claim were presented in 
\cite{hall2001nber}. Application of time series methods for examining 
technology development was presented in details in work~\cite{dernis2000using}.
In this report methods of time series analysis are used to predict the 
development of quantum information processing technologies. 

The analysis was performed on 523 patents selected from the Patent\-Scope
database. After analysing IPC categories, the number of patents in institutions
and the geographic scope of the patents one can draw the following conclusions.

\subsubsection{Patents classification}
\begin{figure}
        \centering
    \includegraphics{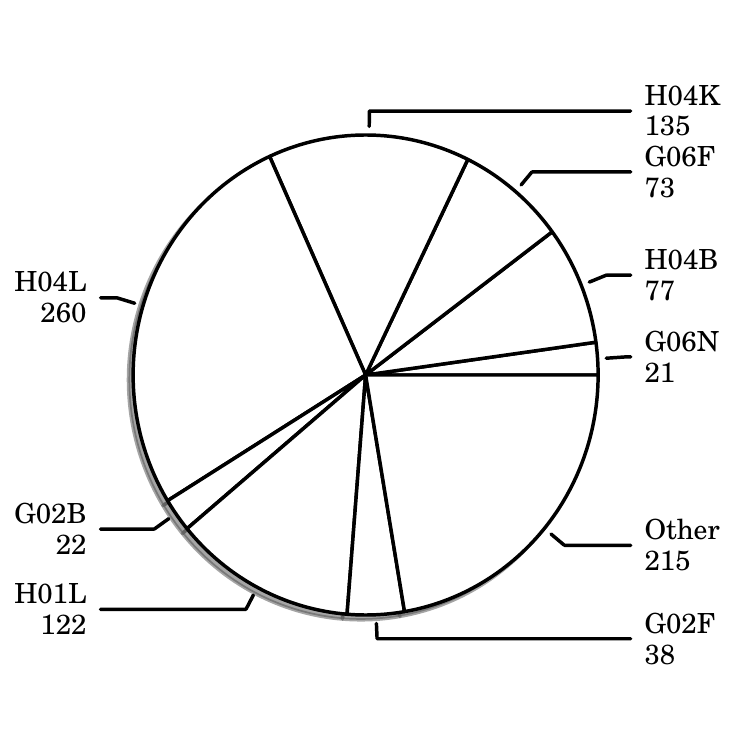}
    \caption{Groups of patents according to the IPC classification. One patent can be included into more than one categories.}
        \label{fig:ipc}
\end{figure}

In Fig.~\ref{fig:ipc} we have gathered the numbers of patents belonging to 
classes of the IPC classification. Relevant IPC codes are listed in 
Tab.~\ref{tab:ipc}. Note that one patent may belong to several classes. Most of
the patents belong to class H04L covering ``Transmission of digital 
information'', two other dominant classes are H04K --- ``Secret communication;
jamming of communication'' and H01L  --- ``Semiconductor devices; electric solid
state devices not otherwise provided for''.

\begin{table}[h]
        \centering
        \begin{tabular}{|p{0.2\textwidth}|p{0.7\textwidth}|}
    \hline
        G02B & Optical Elements, Systems, Or Apparatus\\\hline
                G06F & Electric Digital Data Processing \\\hline
        H04B & Transmission \\\hline
        H04K & Secret Communication; Jamming Of Communication\\\hline
        H04L & Transmission Of Digital Information, E.G. Telegraphic Communication\\\hline
        H01L & Semiconductor Devices; Electric Solid State Devices Not Otherwise
        Provided For\\\hline
        G06N & Computer Systems Based On Specific Computational Models\\\hline
        G02F & Devices Or Arrangements, The Optical Operation Of Which Is
            Modified By Changing The Optical Properties Of The Medium Of The Devices
            Or Arrangements For The Control Of The Intensity, Colour, Phase,
            Polarisation Or Direction Of Light, E.G. Switching, Gating, Modulating
            Or Demodulating; Techniques Or Procedures For The Operation Thereof;
            Frequency-Changing; Non-Linear Optics; Optical Logic Elements; Optical
            Analogue/Digital Converters\\\hline
        \end{tabular}
        \caption{IPC Codes.}
        \label{tab:ipc}
\end{table}
\subsubsection{Applicants}
The number of patents owned by patent holder is presented in
Tab.~\ref{tab:patents-by-applicant}. Its analysis shows that intellectual
property in the field of quantum information processing is highly scattered.
Almost half of the patents are owned by business organizations or research
institutes that have filed up to 5 patents each. Two major firms which are
focused on the area of quantum information  processing, namely MagiQ
Technologies Inc. and D-Wave Systems, own more than 100  quantum information
patents.

\begin{table}
        \centering
        \begin{tabular}{|p{5cm}|r|r|}
        \hline
        Company & No of patents & Percentage \\ \hline
        D-wave & 57 & 11.0\% \\ \hline 
        Magiq Technologies Inc & 52 & 10.0\% \\ \hline 
        Hewlett-packard & 31 & 6.0\% \\ \hline 
        Qinetiq Limited & 25 & 5.0\% \\ \hline 
        Mimos Berhad & 24 & 5.0\% \\ \hline 
        British Telecommunications & 19 & 4.0\% \\ \hline 
        Electronics And Telecommunications Research Institute & 13 & 2.0\% \\ \hline 
        Unisearch Limited & 13 & 2.0\% \\ \hline 
        Japan Science And Technology & 10 & 2.0\% \\ \hline 
        Mitsubishi Denki Kabushiki Kaisha & 10 & 2.0\% \\ \hline 
        The Johns Hopkins University & 9 & 2.0\% \\ \hline 
        Northrop Grumman Systems Corporation & 9 & 2.0\% \\ \hline 
        The Chinese University Of Hong Kong & 9 & 2.0\% \\ \hline 
        Lucent Technologies Inc & 7 & 1.0\% \\ \hline 
        Nec Corporation & 7 & 1.0\% \\ \hline 
        Thales & 6 & 1.0\% \\ \hline 
        Qucor Pty Ltd & 5 & 1.0\% \\ \hline 
        Other & 217 & 41.0\% \\ \hline 
        \end{tabular}
        
    \caption{The number of patents owned by corporations and research institutes.}
    \label{tab:patents-by-applicant}
\end{table}

\subsubsection{Patents by region}
In Tab.~\ref{tab:patents-by-region} we present the numbers of patents filed  in
each of the regions. More than half of the patents were international 
applications filed under the Patent Cooperation Treaty (marked as region WO),
163 patents were filed with the European Patent Office and 51 patents were filed
in  the Republic of Korea.

\begin{table}
        \centering
		\begin{tabular}{|c|l|r|}
		\hline
		Code & Region & Number of patents \\ \hline
		WO & World Intellectual Property Organization & 292 \\ \hline 
		EP & European Patent Office  & 163 \\ \hline 
		KR & Republic of Korea & 51 \\ \hline 
		RU & Russian Federation & 5 \\ \hline 
		ZA & South Africa & 3 \\ \hline 
		IL & Israel & 3 \\ \hline 
		ES & Spain & 3 \\ \hline 
		SG & Singapore  & 2 \\ \hline 
		MX & Mexico & 1 \\ \hline 
		\end{tabular}
        \caption{The number of patents by region.}
    \label{tab:patents-by-region}
\end{table}

\subsubsection{The Nokia case}
It can be observed that there is a trend to extend patent claims so that
embodiments of a method include implementation using quantum information. As an
example Nokia includes such claims in their inventions \textit{e.g} image
analysis. For example in patent \cite{muninder2011method} one can read as
follows:
\begin{quote}
``Computer system 900 is programmed (\textit{e.g.}, via computer program code or
instructions) to detect a face portion in a frame of a plurality of frames in a
multimedia content of a device, track the face portion and perform
color-tracking on losing a track of the face portion for re-tracking the face
portion, as described herein and includes a communication mechanism such as a
bus 910 for passing information between other internal and external components
of the computer system 900. Information (also called data) is represented as a
physical expression of a measurable phenomenon, typically electric voltages, but
including, in other embodiments, such phenomena as magnetic, electromagnetic,
pressure, chemical, biological, molecular, atomic, sub-atomic and quantum
interactions. For example, north and south magnetic fields, or a zero and
non-zero electric voltage, represent two states (0, 1) of a binary digit (bit).
Other phenomena can represent digits of a higher base. \emph{A superposition of
multiple simultaneous quantum states before measurement represents a quantum bit
(qubit).''}
\end{quote}
The last quoted sentence can be found in 257 patents filled by Nokia 
Corporation and in several patents filed by other applicants. In patents filed
in the year 2009 this sentence can be found in three documents, in the year 2010
in 54, in  the year 2011 in 102 and in the year 2012 in 111. It seems that in
2009 organizations started to believe that the realization of a quantum 
computing system
was actually possible.

\section{Models and trends} 

\subsection{Models of time series for the number of filed patents by year}
For the analysis only the data from years 1993-2011 was taken into  account.
Data from year 2012 was rejected due to its incompleteness.  In order to
model the number of filed patents by year, four types of time series models, 
described below, were applied. A similar analysis for patents covering the area
of  biotechnologies was presented in work \cite{jun2010technology}. 

\subsubsection{Linear regression} 
Linear regression finds linear dependency between a pair of variables. The
linear regression model is in the following form:

\begin{equation}
y_i = \alpha + \beta x_i + e_i,
\end{equation}
where
$x_i$ and $y_i$  are dependent variables,
$\alpha$ and $\beta$ are free coefficient and slope coefficient, respectively,
$e_i$ is model error having $0$ mean and variance~$\sigma$.

\subsubsection{Poisson regression}
Poisson regression is a form of regression analysis used to model observations
which can take only non-negative integer values, \textit{e.g.} count data.
It is assumed, that the distribution of the response variable $Y$ is given
by Poisson distribution:
\begin{equation}
P(Y = k) = e^{-\lambda} \frac{\lambda^k}{k!},
\end{equation}
for a positive parameter $\lambda$. Poisson regression is often called the
log-linear model.

\subsubsection{Auto-regression Moving Average models}
Autoregressive moving average (ARMA) models are fitted to time series data
either to forecast future points in the series or to better understand the data.
ARMA models constitute one of the most general classes of time-series
forecasting models, they are applied in cases where data shows stationarity. A
detailed description of this model can be found in the
book~\cite{brockwell2009time}.

\subsubsection{Support Vector Regression model}
The idea of Support Vector Regression (SVR) is based on the designation of the
linear regression function in a space of higher dimension than the original
data.  Mapping to the high-dimensional space is performed by means of non-linear
functions. The SVR method has been successfully used in various fields, such as
time series and forecasting of financial data, approximations of solutions of
complex engineering analyses, convex and quadratic programming, etc.  The theory
for this method was originally developed by V.~Vapnik and his colleagues in
laboratories AT\&T
Bell~\cite{vapnik1963pattern,vapnik1964note,vapnik1996support} for the purpose 
of classification.
However, it is possible to use this method for regression
problems~\cite{tsay2005analysis}.

\subsection{Results}
\begin{figure}[!h]
        \centering
    \includegraphics{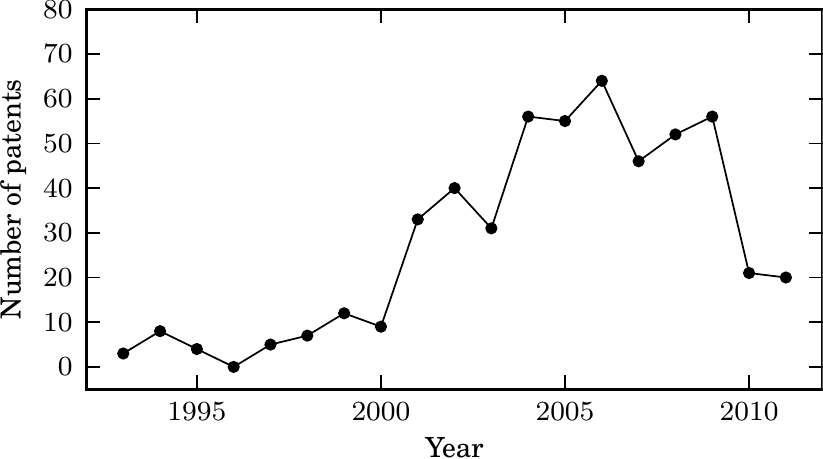}
    \caption{The number of patents in years 1993-2011.}
    \label{fig:patent-data-93-11}
\end{figure}

Figure~\ref{fig:patent-data-93-11} presents the number of reported patents in
years from 1993 to 2011. The above mentioned models were fitted to the data and 
an extrapolation was made to the year 2015.

\subsubsection{Fitting of the models} 

The first step in the analysis was to perform linear regression --- thus the
intercept and slope parameters were determined. The root mean square error RMSE
and coefficient of determination $R^2$ were determined. The results of the
analysis are presented in Table~\ref{tab:reg-lin}. A graphical representation of
the regression model fitted to the data is presented in
Figure~\ref{fig:pred-lin}.

The next step was to perform an analysis of the Poisson regression --- the
results are  presented in Table~\ref{tab:reg-Pois}. The AIC column stands for
the \emph{Akaike factor}, derived from the \emph{Akaike Information Criterion}
and can serve as a measure of quality of the fitting for the Poisson regression.
A graphical representation of the Poisson regression model fitted to the
empirical data is presented in Figure~\ref{fig:pred-pois}.

\begin{table}[h!]
\centering
        \begin{tabular}{|c|c|c|}
                \hline
                $\alpha$ & $\beta$ & $R^2$ \\ \hline
                2.78 & $-5.55\cdot 10^3$ & 0.8391 \\ \hline
        \end{tabular}
        \caption{Results for linear regression}
\label{tab:reg-lin}
\end{table}

\begin{table}[h!] 
\centering
        \begin{tabular}{|c|c|c|}
                \hline
                $\alpha$ & $\beta$ & AIC \\ \hline
                $1.08\cdot 10^{-1}$ & $-2.14\cdot 10^2$ & 283.57 \\ \hline
        \end{tabular}
        \caption{Results for Poisson regression}
\label{tab:reg-Pois}
\end{table}

In the next step SVR models were considered, for which RBF (called the
\emph{radial basis function}) with $\gamma = 1$ was used as the kernel function.
Fitting of the described model is presented in Figure ~\ref{fig:pred-SVR}.

In the last step ARMA models were considered for different model parameters. The
fitting of the ARMA models to the patent data was also presented graphically in
Figure~\ref{fig:pred2}.

On analysing the RMSE ratio, it can be seen that the best fit is obtained for
SVR models. The summary of RMSE values for the considered models is presented in
Table~\ref{tab:RMSE}.

\begin{table}[b!]
        \centering
\begin{tabular}{|c|r|}
	    \hline
	    Model & RMSE \\ \hline
		Poisson regression & 17.81 \\ \hline 
		Linear regression & 15.1 \\ \hline 
		SVR & 7.4 \\ \hline 
		ARMA(0, 1) & 12.44 \\ \hline 
		ARMA(0, 2) & 12.42 \\ \hline 
		ARMA(1, 0) & 12.03 \\ \hline 
		ARMA(1, 1) & 12.03 \\ \hline 
		ARMA(1, 2) & 9.88 \\ \hline 
		ARMA(2, 0) & 12.03 \\ \hline 
		ARMA(2, 1) & 10.53 \\ \hline 
		ARMA(2, 2) & 11.15 \\ \hline 
		\end{tabular}
        \caption{Comparison of the quality of fitting for different models}
        \label{tab:RMSE}
\end{table}

\subsubsection{Forecast based on the introduced models}
In Figures~\ref{fig:pred1} and \ref{fig:pred2} forecast of the number of filed
patents in the years 2012--2015 based on the aforementioned models is presented.
The model which has the best fit, i.e. the SVR model, projects stabilization of
the number of patents per year at the level of around 20. The linear model
forecasts that in the year 2015 approximately 65 patents will be filed. The
Poisson regression model envisages an exponential increase in the number of
patents, however, the estimated values are unlikely because of high RMSE --- the
highest among the used models.  All fitted ARMA models predict an increase in
the number of filed patents. The ARMA(1,2)  model, which has the lowest RMSE
among ARMA models, predicts a slight decrees  and a subsequent increase after
the year 2013.

\newlength{\wdth}
\begin{figure}[!h]
        \setlength{\wdth}{0.45\textwidth}
        \centering
        \subfigure{\includegraphics[width=\wdth]{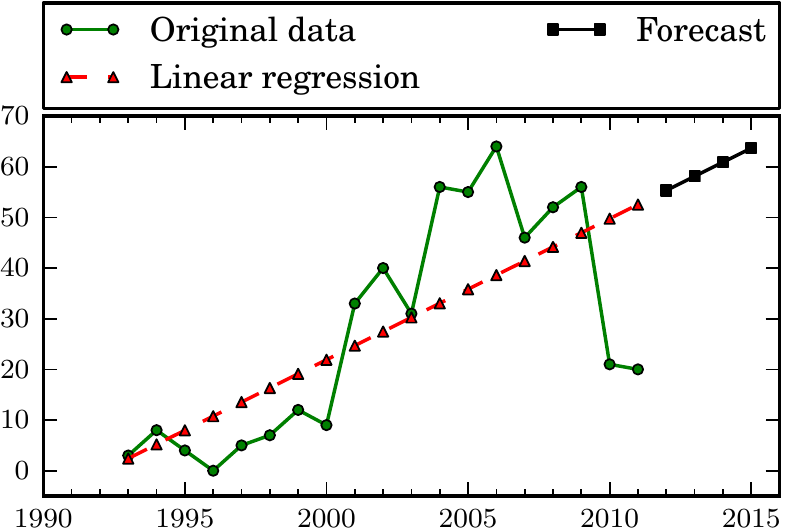}
        \label{fig:pred-lin}
        }
        \subfigure{\includegraphics[width=\wdth]{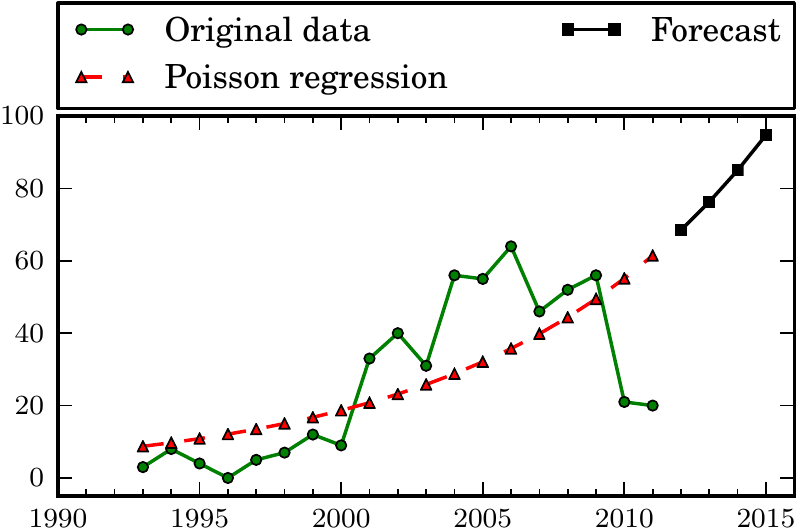}
        \label{fig:pred-pois}
        }\\
        \subfigure{\includegraphics[width=\wdth]{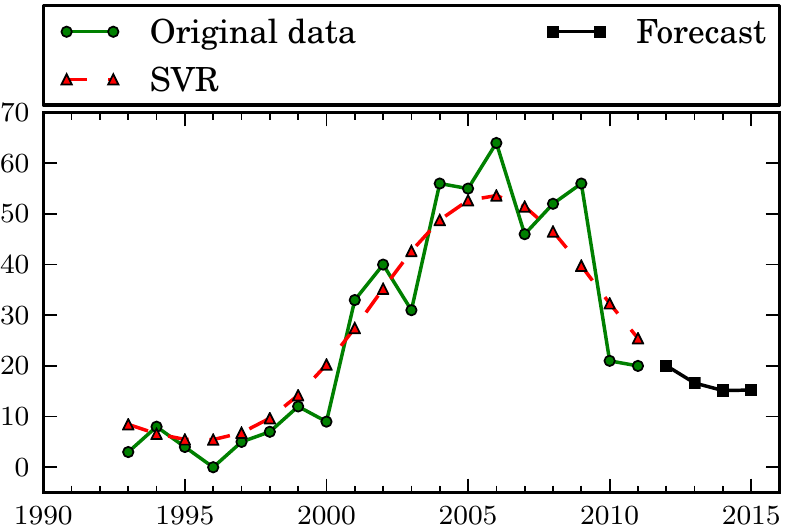}
        \label{fig:pred-SVR}
        }
        \caption{Model fitting and forecast of the number of patents filed in a given year on the basis of linear regression, Poisson regression, and support vector regression.}
        \label{fig:pred1}
\end{figure}

\begin{figure}[!h]
        \setlength{\wdth}{0.45\textwidth}
        \centering
        \subfigure{\includegraphics[width=\wdth]{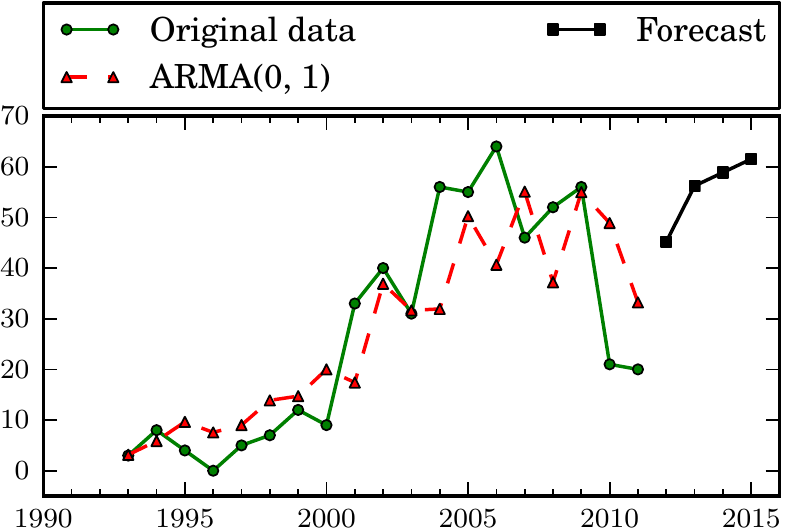}
        \label{fig:pred-ARMA01}
        }
        \subfigure{\includegraphics[width=\wdth]{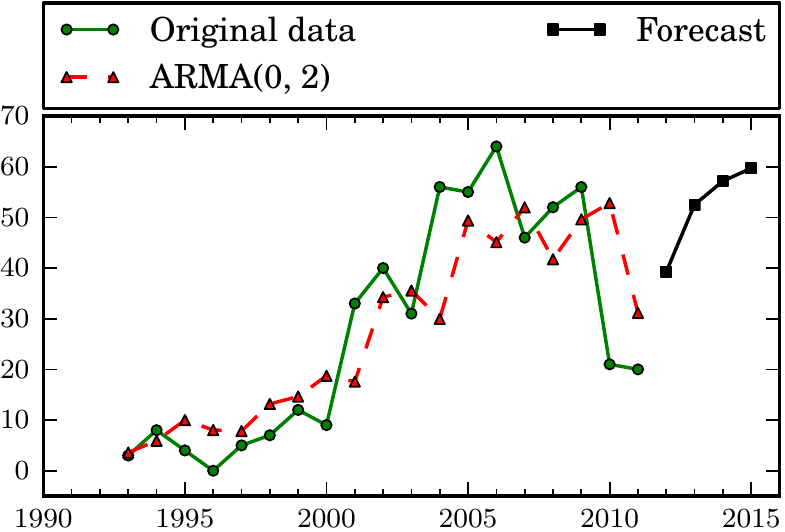}
        \label{fig:pred-ARMA02}
        }\\
        \subfigure{\includegraphics[width=\wdth]{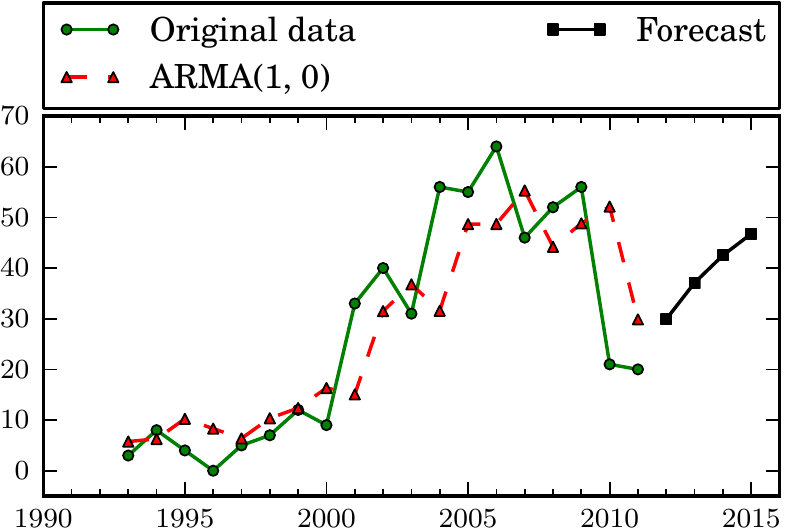}
        \label{fig:pred-ARMA10}
        }
        \subfigure{\includegraphics[width=\wdth]{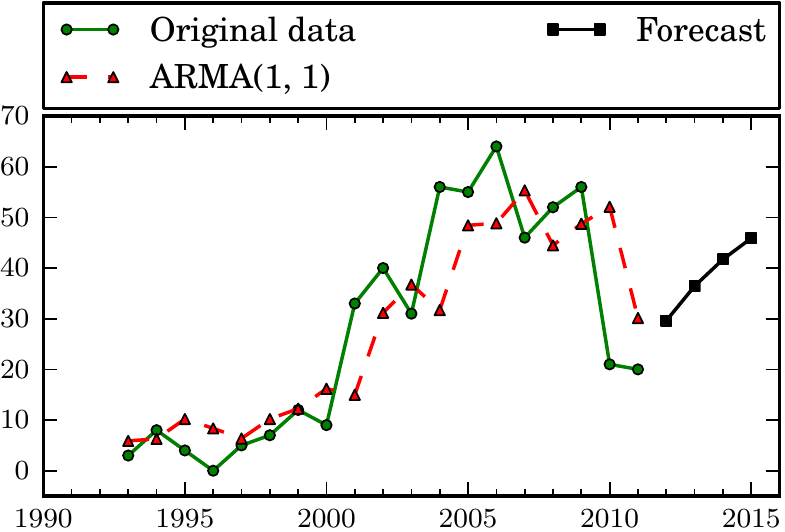}
        \label{fig:pred-ARMA11}
        }\\
        \subfigure{\includegraphics[width=\wdth]{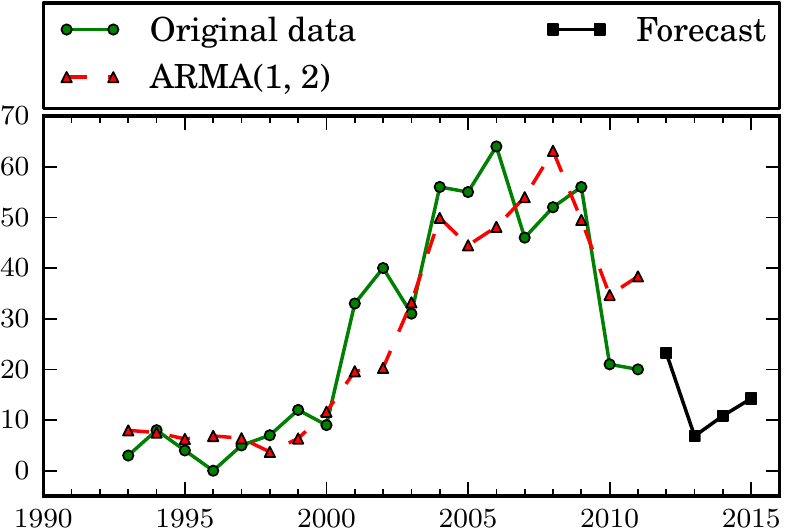}
        \label{fig:pred-ARMA12}
        }
        \subfigure{\includegraphics[width=\wdth]{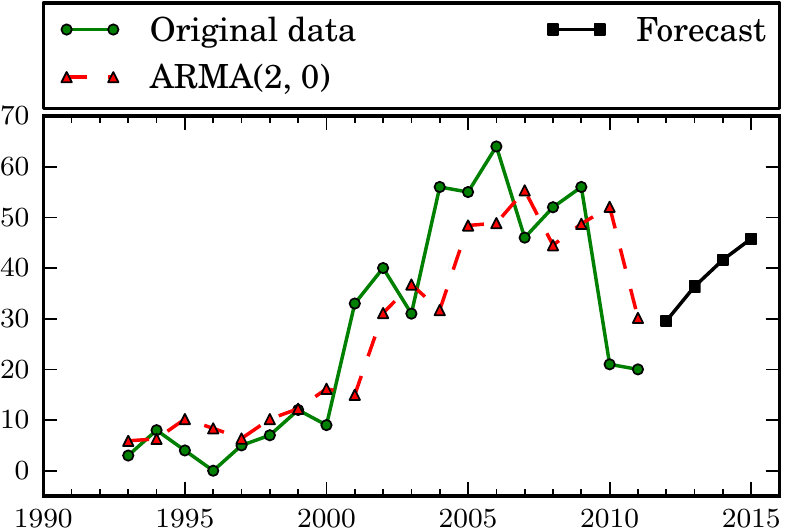}
        \label{fig:pred-ARMA20}
        }\\
        \subfigure{\includegraphics[width=\wdth]{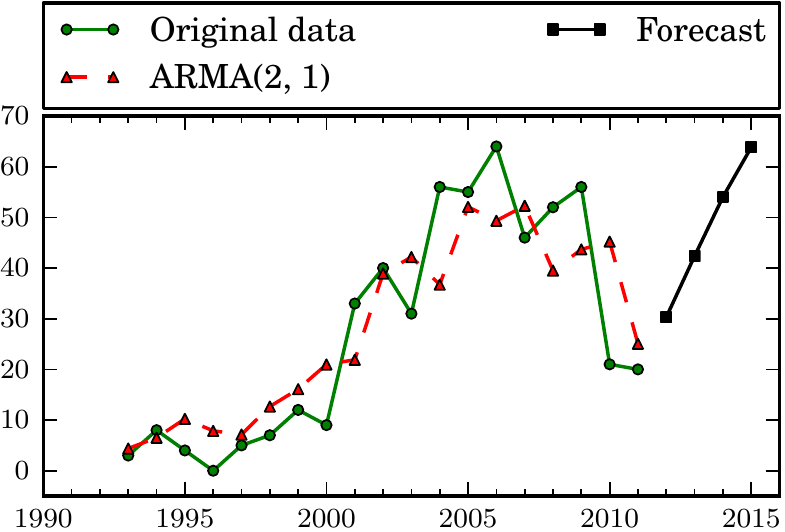}
        \label{fig:pred-ARMA21}
        }
        \subfigure{\includegraphics[width=\wdth]{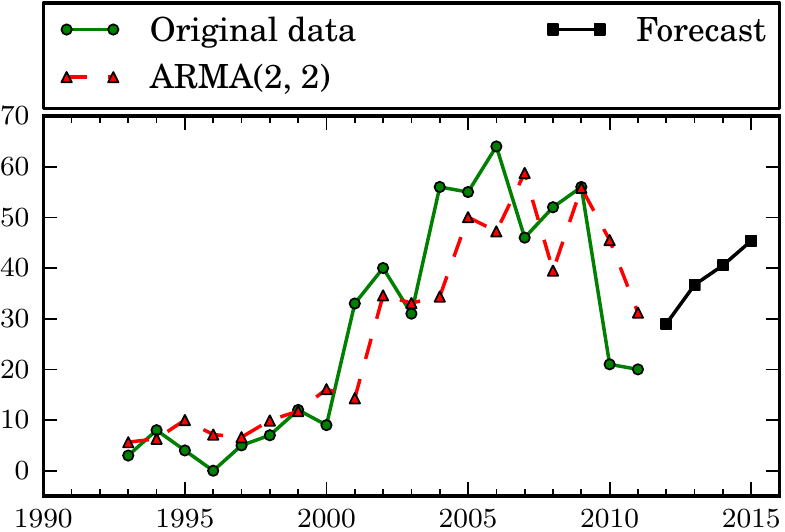}
        \label{fig:pred-ARMA22}
        }\\
        \caption{Model fitting and forecast of the number of patents filed in a given year on the basis of ARMA models.}
        \label{fig:pred2}
\end{figure}

\section{Summary}
We have provided an analysis of patent activity in the field of quantum
information processing. One can notice an increase in patent activity in the 
years 2001-2009. In subsequent years the activity sharply decreased. In order to
predict the future activity we have used several models to estimate the number
of patents to be filed until the year 2015. Both models with low error ---
SVR and ARMA(1,2) --- predict that around 15 patents will be filed in 2015.

\section*{Acknowledgements}
We acknowledge the financial support of the Polish Ministry of Science and
Higher Education under the grant number N N519 442339 and of the research
project No.~WND-POIG.01.01.01-00-021/09: ``Scenarios and development trends of
selected information society technologies until 2025'' funded by the ERDF within
the Innovative Economy Operational Programme, 2006-2013.


\begin{thebibliography}{10}

\bibitem{von1927thermodynamik}
J.~Von~Neumann.
\newblock Thermodynamik quantenmechanischer gesamtheiten.
\newblock {\em G{\"o}ttinger Nachrichten}, 1(11):273--291, 1927.

\bibitem{holevo73information}
Information-theoretic aspects of quantum measurement.
\newblock {\em Probl. Inform. Transm.}, 9(2):31--42, 1973.
\newblock in Russian.

\bibitem{ingarden76quantum}
R.S. Ingarden.
\newblock Quantum information theory.
\newblock 10(1):43--72, 1976.

\bibitem{bennett1984quantum}
C.H. Bennett, G.~Brassard, et~al.
\newblock Quantum cryptography: Public key distribution and coin tossing.
\newblock In {\em Proceedings of IEEE International Conference on Computers,
  Systems and Signal Processing}, volume 175. Bangalore, India, 1984.

\bibitem{ekert1991quantum}
A.K. Ekert.
\newblock Quantum cryptography based on bell’s theorem.
\newblock {\em Physical review letters}, 67(6):661--663, 1991.

\bibitem{feynman1982simulating}
R.P. Feynman.
\newblock Simulating physics with computers.
\newblock {\em International journal of theoretical physics}, 21(6):467--488,
  1982.

\bibitem{deutsch1985quantum}
D.~Deutsch.
\newblock Quantum theory, the church-turing principle and the universal quantum
  computer.
\newblock {\em Proceedings of the Royal Society of London. A. Mathematical and
  Physical Sciences}, 400(1818):97--117, 1985.

\bibitem{shor1994algorithms}
P.W. Shor.
\newblock Algorithms for quantum computation: discrete logarithms and
  factoring.
\newblock In {\em Foundations of Computer Science, 1994 Proceedings., 35th
  Annual Symposium on}, pages 124--134. IEEE, 1994.

\bibitem{grover1996fast}
L.K. Grover.
\newblock A fast quantum mechanical algorithm for database search.
\newblock In {\em Proceedings of the twenty-eighth annual ACM symposium on
  Theory of computing}, pages 212--219. ACM, 1996.

\bibitem{meyer1999quantum}
D.A. Meyer.
\newblock Quantum strategies.
\newblock {\em Physical Review Letters}, 82(5):1052--1055, 1999.

\bibitem{eisert1999quantum}
J.~Eisert, M.~Wilkens, and M.~Lewenstein.
\newblock Quantum games and quantum strategies.
\newblock {\em Physical Review Letters}, 83(15):3077--3080, 1999.

\bibitem{piotrowski2002quantum}
E.W. Piotrowski and J.~S{\l}adkowski.
\newblock Quantum market games.
\newblock {\em Physica A: Statistical Mechanics and its Applications},
  312(1):208--216, 2002.

\bibitem{hemmer09where}
P.~Hemmer and J.~Wrachtrup.
\newblock Where is my quantum computer?
\newblock {\em Science}, 324(5926):473--474, 2009.

\bibitem{walther06cavity}
H.~Walther, B.T.H. Varcoe, B.-G. Englert, and T.~Becker.
\newblock Cavity quantum electrodynamics.
\newblock {\em Reports on Progress in Physics}, 69(5):1325, 2006.

\bibitem{miller05trapped}
R.~Miller, T.E. Northup, K.M. Birnbaum, A.~Boca, A.D. Boozer, and H.J. Kimble.
\newblock Trapped atoms in cavity qed: coupling quantized light and matter.
\newblock {\em Journal of Physics B: Atomic, Molecular and Optical Physics},
  38(9):S551, 2005.

\bibitem{stute12toward}
A.~Stute, B.~Casabone, B.~Brandstätter, D.~Habicher, H.G. Barros, P.O.
  Schmidt, T.E. Northup, and R.~Blatt.
\newblock Toward an ion–photon quantum interface in an optical cavity.
\newblock {\em Applied Physics B}, 107:1145--1157, 2012.

\bibitem{kim09integrated}
J.~Kim and Ch. Kim.
\newblock Integrated optical approach to trapped ion quantum computation.
\newblock {\em Quantum Information \& Computation}, 9:0181--0202, 2009.

\bibitem{montz11-14qubit}
T.~Monz, P.~Schindler, J.T. Barreiro, M.~Chwalla, D.~Nigg, W.A. Coish,
  M.~Harlander, W.~H\"ansel, M.~Hennrich, and R.~Blatt.
\newblock 14-qubit entanglement: Creation and coherence.
\newblock {\em Phys. Rev. Lett.}, 106:130506, 2011.

\bibitem{brennen99quantum}
G.K. Brennen, C.M. Caves, P.S. Jessen, and I.H. Deutsch.
\newblock Quantum logic gates in optical lattices.
\newblock {\em Phys. Rev. Lett.}, 82:1060--1063, 1999.

\bibitem{li08real-time}
T.C. Li, H.~Kelkar, D.~Medellin, and M.G. Raizen.
\newblock Real-time control of the periodicity of a standing wave: an optical
  accordion.
\newblock {\em Opt. Express}, 16(8):5465--5470, 2008.

\bibitem{everitt11creating}
M.S. Everitt, M.L. Jones, B.T.H. Varcoe, and J.A. Dunningham.
\newblock Creating and observing $n$-partite entanglement with atoms.
\newblock {\em Journal of Physics B: Atomic, Molecular and Optical Physics},
  44(3):035504, 2011.

\bibitem{gurudev-dutt07quantum}
M.V. Gurudev~Dutt, L.~Childress, L.~Jiang, E.~Togan, J.~Maze, F.~Jelezko, A.S
  Zibrov, P.R. Hemmer, and M.D. Lukin.
\newblock Quantum register based on individual electronic and nuclear spin
  qubits in diamond.
\newblock {\em Science}, 316(5829):1312--1316, 2007.

\bibitem{maurer12room-temperature}
P.~C. Maurer, G.~Kucsko, C.~Latta, L.~Jiang, N.~Y. Yao, S.~D. Bennett,
  F.~Pastawski, D.~Hunger, N.~Chisholm, M.~Markham, D.~J. Twitchen, J.~I.
  Cirac, and M.~D. Lukin.
\newblock Room-temperature quantum bit memory exceeding one second.
\newblock {\em Science}, 336(6086):1283--1286, 2012.

\bibitem{jordan04entanglement}
A.N. Jordan and M.~B\"uttiker.
\newblock Entanglement energetics in the ground state.
\newblock {\em Journal of Modern Optics}, 51(16-18):2405--2414, 2004.

\bibitem{chirolli06full}
L.~Chirolli and G.~Burkard.
\newblock Full control of qubit rotations in a voltage-biased superconducting
  flux qubit.
\newblock {\em Phys. Rev. B}, 74:174510, Nov 2006.

\bibitem{houck09life}
A.A. Houck, J.~Koch, M.H. Devoret, S.M. Girvin, and R.J. Schoelkopf.
\newblock Life after charge noise: recent results with transmon qubits.
\newblock {\em Quantum Information Processing}, 8:105--115, 2009.

\bibitem{gronbech-jensen10tomography}
Niels Gr\o{}nbech-Jensen, Jeffrey~E. Marchese, Matteo Cirillo, and James~A.
  Blackburn.
\newblock Tomography and entanglement in coupled josephson junction qubits.
\newblock {\em Phys. Rev. Lett.}, 105:010501, Jun 2010.

\bibitem{rigetti12superconducting}
C.~Rigetti, J.~M. Gambetta, S.~Poletto, B.~L.~T. Plourde, J.~M. Chow, A.~D.
  C\'orcoles, J.~A. Smolin, S.~T. Merkel, J.~R. Rozen, G.~A. Keefe, M.~B.
  Rothwell, M.~B. Ketchen, and M.~Steffen.
\newblock Superconducting qubit in a waveguide cavity with a coherence time
  approaching 0.1 ms.
\newblock {\em Phys. Rev. B}, 86:100506, Sep 2012.

\bibitem{stojanovic12quantum-control}
V.M. Stojanovi\'{c}, A.~Fedorov, A.~Wallraff, and C.~Bruder.
\newblock Quantum-control approach to realizing a toffoli gate in circuit qed.
\newblock {\em Phys. Rev. B}, 85:054504, Feb 2012.

\bibitem{kimble08quantum}
H.~J. Kimble.
\newblock The quantum internet.
\newblock {\em Nature}, 453(7198):1023--1030, 2008.

\bibitem{clausen11storage}
Ch. Clausen, I.~Usmani, F.~Bussi\'eres, N.~Sangouard, M.~Afzelius,
  H.~de~Riedmatten, and N.~Gisin.
\newblock Quantum storage of photonic entanglement in a crystal.
\newblock {\em Nature}, pages 508--511, 2011.

\bibitem{hall11ultrafast}
M.~A. Hall, J.~B. Altepeter, and P.~Kumar.
\newblock Ultrafast switching of photonic entanglement.
\newblock {\em Phys. Rev. Lett.}, 106(5):053901, 2011.

\bibitem{choi11athome}
I.~Choi, R.~J. Young, and P.~D. Townsend.
\newblock Quantum information to the home.
\newblock {\em New Journal of Physics}, 13:063039, 2011.

\bibitem{patentscope}
{PatentScope}.
\newblock http://patentscope.wipo.int/.

\bibitem{chang}
M.C. Chang.
\newblock Quantum computation patent mapping-a strategic view for the
  information technique of tomorrow.
\newblock In {\em Services Systems and Services Management, 2005. Proceedings
  of ICSSSM'05. 2005 International Conference on}, volume~2, pages 1177--1181.
  IEEE, 2005.

\bibitem{hall2001nber}
B.H. Hall, A.B. Jaffe, and M.~Trajtenberg.
\newblock The {NBER} patent citation data file: Lessons, insights and
  methodological tools.
\newblock Technical report, National Bureau of Economic Research, 2001.

\bibitem{dernis2000using}
H.~Dernis and D.~Guellec.
\newblock Using patent counts for cross-country comparisons of technology
  output.
\newblock {\em STI-Science Technology Industry Review}, (27):129--146, 2000.

\bibitem{muninder2011method}
V.~Muninder, P.~Mishra, C.-Y. Jo, and K.~Govindarao.
\newblock {Method, Apparatus and Computer Program Product for Tracking Face
  Portion}.

\bibitem{jun2010technology}
S.~Jun and D.~Uhm.
\newblock Technology forecasting using frequency time series model:
  Bio-technology patent analysis.
\newblock {\em Journal of Modern Mathematics and Statistics}, 4(3):101--104,
  2010.

\bibitem{brockwell2009time}
P.J. Brockwell and R.A. Davis.
\newblock {\em Time series: theory and methods}.
\newblock springer Verlag, 2009.

\bibitem{vapnik1963pattern}
V.~Vapnik.
\newblock Pattern recognition using generalized portrait method.
\newblock {\em Automation and Remote Control}, 24:774--780, 1963.

\bibitem{vapnik1964note}
V.~Vapnik and A.~Chervonenkis.
\newblock A note on one class of perceptrons.
\newblock {\em Automation and Remote Control}, 25(1), 1964.

\bibitem{vapnik1996support}
V.~Vapnik, S.E. Golowich, and A.~Smola.
\newblock Support vector method for function approximation, regression
  estimation, and signal processing.
\newblock In {\em Advances in Neural Information Processing Systems 9}, 1996.

\bibitem{tsay2005analysis}
R.S. Tsay.
\newblock {\em Analysis of financial time series}, volume 543.
\newblock Wiley-Interscience, 2005.

\end{thebibliography}
\end{document}